# MEASURING INTERGROUP AGREEMENT AND DISAGREEMENT

MADHUSMITA PANDA, SHARAYU PARANJPE AND ANIL GORE

Cytel Statistical Software & Services Private Limited, Pune, India

Correspondence should be sent to

Madhusmita Panda,

6th, 7th Floor, Lohia-Jain IT Park – A Wing,

Survey #150, India, Paud Rd, Bhusari Colony,

Kothrud, Pune, Maharashtra 411038

+91 20 6709 0295

Madhsmita.panda@cytel.com

**Abstract**

This work is motivated by the need to assess the degree of agreement between two independent groups of raters. One simple way of measuring intergroup agreement is to use the average value of Cohen's kappa. It can be calculated for each pair of raters (one rater from each group) and then average can be taken over all pairs (pair-wise agreement measure). Alternatively, data on all the raters can be pooled for each group disregarding the individual rater and just one measure can be calculated (pooled agreement measure). Apart from the above intuitive ways, we also propose two new methods to measure agreement between two groups of raters. First is an agreement measure which is the cube root of product of agreement values within each group and in the combined group (geometric mean). Second is disagreement measure which is a quadratic form of difference vectors. Properties of all four measures are investigated.

# 1. Introduction

In clinical studies, comparison of a new measurement technique with an established one is often needed to see whether they agree sufficiently for the new to replace the old. Such comparisons are sometimes done inappropriately, notably by using correlation coefficients when the data is on a continuous scale. The use of correlation is misleading since highly correlated measures can disagree everywhere unless the relationship between the two is essentially the line X=Y. Hence instead of calculating correlation, a better alternate approach is to visualize the nature of agreement using Bland-Altman plot. The plot is possible even if the measurements are only ranks. Several measures are available in the literature to quantify agreement between two individual raters when ratings are on a categorical scale. They are Percent agreement (Hartmann, 1977), Cohen's kappa coefficient (Cohen, 1960), Intra-class kappa coefficient (Kraemer, 1979) and Weighted kappa coefficient (Cohen, 1968).

When the need is more complex viz. to measure agreement among several raters, Cohen's kappa can be used in a generalized form as Fleiss' kappa (Fleiss, 1981). Another measure of agreement used in this case is Krippendorff's alpha (Krippendorff, 2004). It is applicable to qualitative as well as quantitative data. A further generalization is needed in practice when the agreement between two independent groups of raters is to be assessed. For example, agreement between a group of expert (in house) assessors and a group of lay consumers, when judging the same product, is of interest before a product is marketed. If the agreement is poor, suitability of the expert group can be questioned. Another example arises when two groups of physicians with different affiliations or experience diagnose the same group of patients clinically.

Methods for testing of agreement between two groups of raters, ranking same items, were proposed by Schucany & Frawley (1973), Hollander & Sethuraman (1978), Kraemer (1981) and Feigin & Alvo (1986). These methods are generally based on the Spearman rank correlation coefficient or Kendall's tau coefficient. Agreement is about closeness of ratings given by the raters while correlation is more about the similarity in rating pattern rather than closeness i.e. it measures the strength of a relation between two variables, not the agreement between them. Hence using correlation coefficients for measuring agreement may not be appropriate.

Very few methods are available in literature for estimating the degree of agreement between two groups of raters, when evaluations are on a categorical scale. Schouten (1982) proposed a weighted kappa coefficient. Vanbelle (2009) also proposed an extension of 'Cohen's kappa and weighted Cohen's kappa coefficient for two raters' to two groups of raters. For rating on ordinal scale one way of measuring agreement between two groups is by consensus method. Ratings given by a group for a subject are replaced by a single value called 'consensus value'. There are different ways of defining consensus value e.g. the modal rating (e.g., van Hoeij et al., 2004) or the median rating (e.g., Raine et al., 2004).Consensus approach reduces the problem from two groups to two 'raters' and then any method such as Cohen's kappa can be applied. Mode as a consensus measure can be used even if rating is categorical. Apart from these measures, Percent agreement (Hartmann, 1977) or proportion agreement is another candidate to assess the degree of agreement between two independent groups of raters. Here, for each subject, ratings across groups are compared and cases of agreement are counted. They are then totaled over subjects and divided by the total number of possible comparisons.

The objective of the present work is to develop an overall index of agreement between two independent groups of raters, for continuous as well as categorical scale, taking into account the heterogeneity of each group. This paper proposes four new indices. First two are pairwise and pooled agreement indices calculated using Cohen's kappa (in case of qualitative data). The third index is a cube-root of the product of three intra-group agreement measures (within each group separately and within a union of two groups). In other words, it is the geometric mean of the three indices. The last index is based on a quadratic form to measure the degree of disagreement. This method is useful only when rating is ordinal.

The proposed indices are calculated for an illustrative data set. Their sampling variability and empirical distribution are determined by re-sampling method. Comparison with available indices suggests that the proposed measures may be more precise than currently available measures. Among the four proposed measures, the index of disagreement may have an edge over the other measures.

## 2. Intuitive Measures of Agreement Between Two Groups, based on Cohen's kappa

Cohen's kappa measures the degree of agreement for categorical assessments made by two raters while assessing the same subjects. We can use this measure to calculate agreement between two groups in an intuitive way. There are two possible ways. One is to calculate agreement between each pair of raters [one rater from each group] and then take the average of kappa values over all the pairs. Another way is to pool the ratings of each subject for each group disregarding the individual raters i.e. to take the score given by group 1 raters for each subject in one column and the respective score given by the group 2 raters for the same subject in another column. Now it is easy to treat the two columns as two raters and calculate kappa. Resulting agreement values with the two approaches tend to be rather close.

## 3. Proposed Methods for Intergroup Agreement and Disagreement

The above explained methods include a measure which calculates agreement between two raters and we generalized it to calculate agreement between two groups of raters. This led us to think that instead of using Cohen's kappa, it is possible to use Fleiss' or Krippendorff's alpha to create a new measure. Alternatively, one may directly examine the differences between ratings given to a subject by two raters (one from each group). Difference if any will suggest disagreement. Such differences may be aggregated in a suitable manner. These ideas lead to two new measures. They are discussed in this section.

### 3.1. Cube Root of Product Measure

Consider two groups, group A with $m_1$ raters and group B with $m_2$ raters. Each member of each group examines the same set of n subjects. We can calculate three intragroup agreement values; Agr(A) for the first group of $m_1$ raters, Agr(B) for the second group of $m_2$ raters and Agr(AUB) for a combined group of $m_1+m_2$ raters. Then the proposed cube root of product measure (geometric mean) is given by

$$CRPm = \sqrt[3]{Agr(A) * Agr(B) * Agr(AUB)}$$

The intragroup agreement can be calculated using Fleiss' kappa or Krippendorff's alpha depending upon the data. An intuitive interpretation of the measure is as follows: If the raters in each group are in complete agreement then CRPm = 1. If there is no agreement among the raters other than what would be expected by chance then CRPm ≤ 0. Proposed measure assumes high values only when each of the three factors has high value. In other words, we are saying that intergroup agreement is high only when agreement within each group is high and also agreement in the combined group is high. On the other hand, if there is any heterogeneity in any of the groups it will affect the whole measure i.e. if any one of the three factors is small; the overall measure results in a small value.

### 3.2. Disagreement Measure

Again consider two groups, group A with $m_1$ raters and group B with $m_2$ raters. Each member of each group examines the same set of n subjects. In this method, data is expressed in a multivariate form. Consider a vector with $m_1$ components which are differences between the scores given by the $m_1$ raters of group A and $j^{th}$ rater of group B for $k^{th}$ subject. We will have $m_2*n$ such difference vectors. Let $X_{ijk}$ denote the difference in rating given by $i^{th}$ rater of group A and $j^{th}$ rater of group B on $k^{th}$ subject. Then the vector is defined as

$$\underline{X}_{jk} = [A_{1k} - B_{jk}, A_{2k} - B_{jk}, \ldots, A_{m_1k} - B_{jk}]'$$

Where $j = 1, 2, \ldots, m_2$; $k = 1, 2, \ldots, n$; $A_{ik}$ is the rating of subject k by rater i in group A; $B_{jk}$ being analogous. If there is complete agreement between the two groups of raters then each of these vectors will be null. Departure from zero differences suggests that there is disagreement. Next step is to combine all these differences. The following quadratic form achieves a drastic reduction in dimensions:

$$Q_{jk} = \underline{X}_{jk}' * S^{-1} * \underline{X}_{jk} \quad \text{where S is var-cov matrix of } m_2*n \text{ vectors } \underline{X}_{jk}.$$

Note that the above quadratic form is positive definite i.e. it always exceeds zero (unless all differences are zero). The other property of a quadratic form is $\underline{X}_{jk}' * S^{-1} * \underline{X}_{jk} \leq \lambda_1 \underline{X}_{jk}' * \underline{X}_{jk}$ where $\lambda_1$ is largest eigen root of S. Therefore, we take the average over j and k of the ratio $\frac{X_{jk}' * S^{-1} * X_{jk}}{X_{jk}' * X_{jk}}$ and further divide it by $\lambda_1$. Dividing by $\lambda_1$ will normalize the measure and the measure will lie between 0 and 1. The final formula of the disagreement measure will be

$$Dm = \frac{\left(\sum_{k=1}^{n}\sum_{j=1}^{m_2} \frac{X_{jk}' * S^{-1} * X_{jk}}{X_{jk}' X_{jk}}\right) \Big/ (m_2 * n)}{\lambda_1}$$

The measure can work for continuous valued variables as well as ordinal categorical variables as long as the latter is coded. Secondly, the measure reflects disagreement. Greater its value greater is the disagreement between the two groups. Since our measure lies between 0 and 1 the proposed agreement measure can be given as $PAm = 1 - Dm$.

## 4. Illustrative data

Consider a situation in which 20 subjects are asked to describe in a paragraph their assessment of the winner of recent Presidential election. Table 1 provides hypothetical ratings given by two groups of coders for these descriptions. Each coder has to judge the opinion of the subject and categorize it on a scale of 1 to 5, 1 indicating a strongly adverse assessment and 5 indicating a strongly favorable assessment. The first group is of three professional coders while the second group is of three naive coders. The aim is to measure the agreement between expert coders and naive coders. This dataset is used to calculate values of different indices and also to study the behavior of indices using a re-sampling approach.

When we have two raters and we want to measure agreement between them, the first step is to plot the data and draw the line of equality (X=Y) on which all points should lie if the two raters gave exactly the same reading every time. Similarly, we will plot the data (from Table 1) of the two groups for which we want to estimate agreement. Our objective is to find whether these two groups agree while coding the description given by the subjects.

The graph will be plotted as coding given by expert coders (EC) vs. coding given by naïve coders (NC) and if the points lie on the line of equality then we can say the raters of these two groups gave exactly the same rating every time i.e. they are in agreement. This type of plot gives some idea regarding the agreement between the raters/ groups of raters.

There were total 20*3*3 = 180 data points (total possible pairs for all subjects from Table 1). Note that the number of distinct values is small and each point in the graph in Figure 1 represents many cases. The graph shows that there are values which lie on the line of equality. But there are other values which are either above or below the line of equality. From Table 1, we can see that more than 72% observations (i.e 130/180) lie on the line of equality. Hence we can say that there seems to be good agreement between the two groups. To take into account the cases with disagreement together with the extent of difference in rating we need a more elaborate measure. Lastly, we should examine the sampling variability of the measure to be used. This is described in the next section.

## 5. Sampling variance

The Jackknife resampling method (Efron & Tibshirani, 1993) was used to determine the sampling variance of the agreement indices. Suppose we want to estimate agreement between two groups of raters who are assessing the same number of subjects, say n. Let $\hat{\theta}$ denote the value of index of agreement between the two groups of raters. The jackknife estimator of a parameter is found by systematically leaving out data on each subject from a dataset, calculating the estimate each time and then finding the average of these estimates. Given a sample of size n, the jackknife estimate is found by aggregating the n estimates based on n-1 values in each subsample. Let $\hat{\theta}_i$ denote the agreement index leaving out the data on $i^{th}$ subject. The Jackknife estimator of the agreement index is then defined by

$$\bar{\theta}_{(.)} = \frac{1}{n} \sum_{i=1}^{n} \hat{\theta}_i$$

with variance

$$Var_{jackknife} = \frac{n-1}{n} \sum_{i=1}^{n} (\hat{\theta}_i - \bar{\theta}_{(.)})^2$$

The jackknife technique can be used to estimate the bias of an estimator calculated over the entire sample. Say $\hat{\theta}$ is the calculated estimator of the parameter of interest based on all $n$ observations. The jackknife estimate of the bias of $\hat{\theta}$ is given by:

$$\widehat{Bias}_\theta = (n-1)(\bar{\theta}_{(.)} - \hat{\theta})$$

and the resulting bias-corrected jackknife estimate of θ is given by:

$$\hat{\theta}_{jack} = n\hat{\theta} - (n-1)\bar{\theta}_{(.)}$$

Efron (1979) suggested another way to think about the jackknife in terms of the pseudo-values. The i[th] pseudo-value is given by

$$ps_i = n\hat{\theta} - (n-1)\hat{\theta}_i$$

The basic jackknife recipe is to treat the pseudo-values $ps_i(X)$ as if they were independent random variables with mean θ. One can then obtain confidence interval for θ and carry out statistical tests about θ using the Central Limit Theorem. Now, the mean and variance of the pseudo-values are

$$\overline{ps} = \frac{1}{n}\sum_{i=1}^{n} ps_i \quad and \quad Var(ps_i) = \frac{1}{(n-1)}\sum_{i=1}^{n}(ps_i - \overline{ps})^2$$

and the jackknife 95% confidence interval for θ is

$$\left(\overline{ps} - 1.96\sqrt{\frac{1}{n}Var(ps_i)}, \overline{ps} + 1.96\sqrt{\frac{1}{n}Var(ps_i)}\right)$$

Similarly, one can define a jackknife p-value for the hypothesis $H_0: \theta = \theta_0$ by comparing

$$Z = \frac{\overline{ps} - \theta_0}{\sqrt{\frac{Var(ps_i)}{n}}}$$

with a standard normal variable.

Mean and standard deviation of proposed measures and other four measures from the literature are estimated for the data in Table 1 using the jackknife pseudo-value method. From the results given in Table 2, notice that the standard errors of proposed agreement measures are smaller than the standard errors of available measures. Consensus measure has the highest standard error. Comparison of mean values of different measures for the same data may not be as useful since different measures refer to very different entities.

The data provided in the Table 1 is ordinal. Hence, Vanbelle's generalized linear weighted kappa method is used. Similarly for calculating CRPm, Krippendorff's alpha is used and linear weighted Cohen's kappa is used for Pairwise and Pooled agreement measure.

Note that the proposed agreement measure (1-Dm) has the lowest and Consensus (mode) measure has the highest standard error. We further investigate the empirical distribution of these measures in next section.

## 6. Empirical Distribution

Empirical distributions of the eight contending measures of agreement are obtained for the data in Table 1 using bootstrap approach. Bootstrap is a statistical technique popularized by Bradley Efron in 1979. Bootstrap method draws samples with replacement from the empirical distribution of data and computes any statistic T to obtain its sampling distribution. 95% confidence interval for the population parameter is given by 2.5% and 97.5% quantiles of the sampling distribution.

With the data from Table 1, 1000 bootstrap samples were generated and histograms were plotted for all the eight measures. From the histograms (see Figure 2), we can see that except PAm (1-Dm) and Consensus measure, the empirical distributions of all the other measures are relatively symmetric. The distribution of Proposed Agreement measure (PAm) is left skewed.

## 7. Discussion

The problem of assessing agreement between two groups of raters is often faced in practice but in the literature there are just a few measures available for the same. We have proposed several new measures to estimate the intergroup agreement and have examined their performance relative to Proportion agreement measure, Vanbelle's generalized kappa measure and consensus measure. Quadratic form based measure (PAm) seems better than available measure in terms of (resampling based) standard error and confidence interval. Though PAm has the smallest standard error, its empirical distribution is not symmetric. Further, a word of caution is in order. Since it is based on the inverse of a covariance matrix, singularity of the matrix can be problematic. It can arise if two raters in the same group are in very close agreement. In such a case, it may be convenient to drop one of these two raters and carry out the analysis as usual. It may be argued that dropping one rater will result in underestimating agreement. In that case, a generalized inverse of covariance matrix can be considered.

Proportion agreement measure has a low standard error and its distribution looks symmetric. But proportion agreement is based on the number of agreement cases and it neglects the cases where there is a disagreement between the raters. The measure treats the ordinal data same as nominal data. Also, Cohen (1960) and Hunt (1986) has suggested that the use of Proportion (or percent) agreement to measure inter-rater agreement should be discouraged, because it does not take into account the agreement due solely by chance. It tends to overestimate agreement and hence can be misleading. But, while the actual value of the measure may have limitations in use, it may be relevant for comparison of different situations.

Next, Vanbelle's generalized linear weighted kappa has the third lowest standard error. Its empirical distribution is also approximately symmetric but it is more biased as compare to PAm. Consensus (Median) measure has next lowest standard error followed by Pooled and Pairwise agreement measure. But the empirical distribution of Consensus (Median) measure is not symmetric whereas the empirical distribution of Pooled and Pairwise agreement measure is approximately symmetric. Also, in Consensus measure we lose a lot of data. The inputs of all the raters are not considered here and it may be the case that the rating chosen for the group might be given by very few raters. Lastly, CRPm has symmetric empirical distribution. But the reliability of the measure is questionable due to its high standard error.

The methods proposed for measuring agreement between two groups of raters in this paper are useful for all kind of data. The properties of all the proposed measures are examined for rank data but their examination may also be of interest for quantitative data which will need further investigation.

*Acknowledgement*: - We would like to thank Cytel Statistical Software & Services Pvt. Ltd. for sponsoring this research work.

Measuring Intergroup Agreement And Disagreement                                        14# References

## Tables

Table 1 : Scores given by 3 expert coders and 3 naïve coders on ordinal scale for 20 subjects.

| Subject # | EC 1 | EC 2 | EC 3 | NC 1 | NC 2 | NC 3 | #Pairs in Agreement |
|---|---|---|---|---|---|---|---|
| 1 | 1 | 1 | 1 | 2 | 1 | 2 | 3 |
| 2 | 1 | 1 | 1 | 1 | 2 | 1 | 6 |
| 3 | 2 | 3 | 2 | 2 | 3 | 2 | 5 |
| 4 | 1 | 1 | 1 | 2 | 1 | 1 | 6 |
| 5 | 1 | 1 | 1 | 1 | 2 | 1 | 6 |
| 6 | 1 | 1 | 1 | 1 | 1 | 2 | 6 |
| 7 | 2 | 4 | 3 | 3 | 3 | 4 | 3 |
| 8 | 1 | 1 | 1 | 1 | 1 | 1 | 9 |
| 9 | 1 | 1 | 1 | 1 | 1 | 1 | 9 |
| 10 | 1 | 1 | 1 | 1 | 1 | 1 | 9 |
| 11 | 2 | 3 | 2 | 3 | 2 | 3 | 4 |
| 12 | 1 | 1 | 1 | 1 | 1 | 1 | 9 |
| 13 | 1 | 1 | 1 | 1 | 1 | 1 | 9 |
| 14 | 1 | 1 | 1 | 1 | 1 | 1 | 9 |
| 15 | 1 | 2 | 1 | 1 | 2 | 1 | 5 |
| 16 | 1 | 1 | 1 | 1 | 1 | 1 | 9 |
| 17 | 2 | 1 | 2 | 2 | 1 | 2 | 5 |
| 18 | 1 | 1 | 1 | 1 | 1 | 1 | 9 |
| 19 | 5 | 5 | 5 | 4 | 5 | 5 | 6 |
| 20 | 2 | 4 | 3 | 3 | 3 | 4 | 3 |
|  |  |  |  |  |  | Total | 130 |

EC: Expert Coder, NC: Naïve Coder

1: Strongly against  2: Mildly against  3: Neither against nor favorable  4: Mildly favorable  5: Strongly favorable

Table 2 : Performance of 8 different measures of agreement between two groups of raters based on data in Table 1.

| Method | Agreement Value ($\hat{\theta}$) | Jackknife Statistics | | |
|---|---|---|---|---|
| | | Mean ($\overline{ps}$) | SE | CI |
| *Proposed Agreement Measure(1 – Dm) | 0.964 | 0.955 | 0.018 | (0.9198, 0.9897) |
| *Cube Root of Product Measure | 0.777 | 0.807 | 0.108 | (0.5948, 1.0000) |
| *Pairwise Agreement Measure | 0.702 | 0.739 | 0.106 | (0.5305, 0.9474) |
| * Pooled Agreement Measure | 0.706 | 0.741 | 0.101 | (0.5419, 0.9392) |
| +Proportion Agreement Measure | 0.722 | 0.722 | 0.057 | (0.6099, 0.8345) |
| +Vanbelle's Generalized Measure | 0.817 | 0.844 | 0.077 | (0.6930, 0.9960) |
| +Consensus (Median) Measure | 0.891 | 0.913 | 0.091 | (0.7337, 1.0000) |
| +Consensus (Mode) Measure | 0.850 | 0.921 | 0.176 | (0.5773, 1.0000) |

$\overline{ps}$ : Mean of the pseudo-values, SE: Standard Error, CI: Confidence Interval

*Proposed Measures    +Measure available in literature

**Figures**

Figure 1 : Scatter plot between the scores given by the group of Expert coders and Naïve coders.

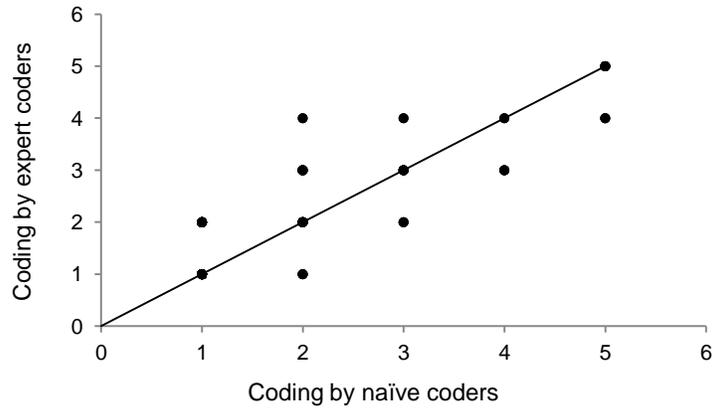

Figure 2 : Histograms representing empirical distribution of the measures.

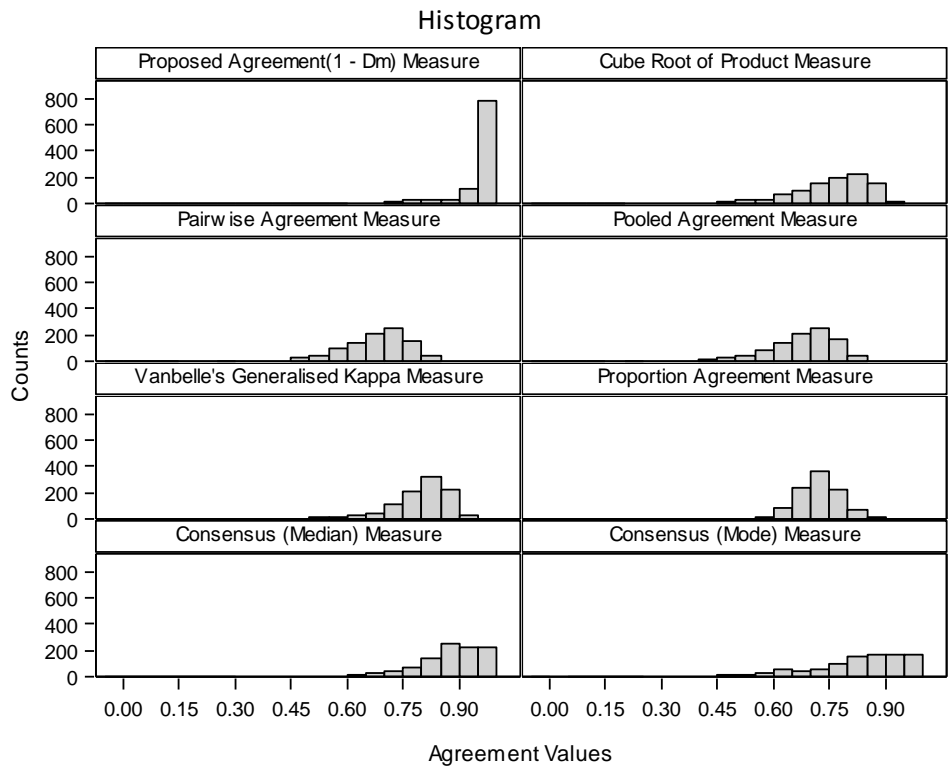